# SHOULD I USE METAVERSE OR NOT? AN INVESTIGATION OF UNIVERSITY STUDENTS' BEHAVIORAL INTENTION TO USE METAEDUCATION TECHNOLOGY

Nikolaos Misirlis[1], Yiannis Nikolaidis[2] and Anna Sabidussi[1]
[1]*HAN University of Applied Sciences, International School of Business, The Netherlands*
[2]*University of Macedonia, School of Applied Informatics, Greece*

**ABSTRACT**

Metaverse, a burgeoning technological trend that combines virtual and augmented reality, provides users with a fully digital environment where they can assume a virtual identity through a digital avatar and interact with others as they were in the real world. Its applications span diverse domains such as economy (with its entry into the cryptocurrency field), finance, social life, working environment, healthcare, real estate, and education. During the COVID-19 and post-COVID-19 era, universities have rapidly adopted e-learning technologies to provide students with online access to learning content and platforms, rendering previous considerations on integrating such technologies or preparing institutional infrastructures virtually obsolete. In light of this context, the present study proposes a framework for analyzing university students' acceptance and intention to use metaverse technologies in education, drawing upon the Technology Acceptance Model (TAM). Specifically, the study aims to investigate the relationship between students' intention to use metaverse technologies in education, hereafter referred to as MetaEducation, and selected TAM constructs, including Attitude (ATT), Perceived Usefulness (PU), Perceived Ease of Use (PE), Self-efficacy (SE) of metaverse technologies in education, and Subjective Norm (SN). Through the development of a structural model of MetaEducation acceptance, the study aims to provide insights to university managers, policymakers, and professors for effectively incorporating this emerging technology into educational settings. Preliminary findings reveal a hesitance among university students to adopt MetaEducation technologies. Notably, Self-efficacy and Subjective Norm have a positive influence on Attitude and Perceived Usefulness, whereas Perceived Ease of Use does not exhibit a strong correlation with Attitude or Perceived Usefulness. The authors postulate that the weak associations between the study's constructs may be attributed to limited knowledge regarding MetaEducation and its potential benefits. Further investigation and analysis of the study's proposed model are warranted to comprehensively understand the complex dynamics involved in the acceptance and utilization of MetaEducation technologies in the realm of higher education.

# SHOULD I USE METAVERSE OR NOT?
# AN INVESTIGATION OF UNIVERSITY STUDENTS' BEHAVIORAL INTENTION TO USE METAEDUCATION TECHNOLOGY

## 1. INTRODUCTION

The concept of Metaverse technology, which combines and includes lifelogging, mirror world, virtual and augmented reality, has been permeating various aspects of daily life for the past 25 years (Kaddoura and Al Husseiny, 2023). It was initially introduced to a wider audience through the cinema industry, and has since evolved to encompass a diverse range of applications. Augmented reality constitutes the majority of Metaverse technology, accounting for approximately 70% of its functionality. The digital environment of the Metaverse encompasses virtual cities, real estate, schools, clubs, restaurants, and more, mirroring the elements of the physical world.

The potential fields of application for Metaverse technology are vast and ever-expanding. From economy and finance (Ko et al., 2021), to social life and work, eHealth (Misirlis et al., 2021), real estate (Terdiman, 2007), and lastly education (Collins, 2008) the possibilities are continuously increasing.

In the realm of education, the integration of digital tools in schools of all levels has become a prevalent trend, further accelerated by the post-COVID-19 era where the functional role of schools has been redefined (Park, 2009). The potential benefits of metaverse use in education range from gamifications to enabling skill-based learning, enhancing diversity and inclusion in educational experiences, and creating a pleasant learning environmental (Kaddoura and Al Husseiny, 2023).

However, the process of incorporating digital tools into education is not without challenges. Issues related to the technological knowledge and infrastructure of educational institutions, the cost of acquiring new equipment, and the readiness of faculty and students to embrace these technologies can all impact the adoption of MetaEducation. Moreover, the acceptance of such cutting-edge technology is still a subject of ongoing research due to its innovative nature.

Various factors can influence the levels of acceptance of MetaEducation. Cultural differences among students from different countries, varying budgets of educational institutions, and divergent perspectives on the future of education can all affect the reception of this technology. Faculty members and students may not always be fully aware of the potential benefits that MetaEducation can bring to their academic lives, which may impact their willingness to embrace and adopt this state-of-the-art technology. The paper is structured as follows. The next chapter presents the objectives and the hypotheses of the research. Second, an in-depth literature review on TAM is presented, followed by the analysis of the chosen methodology. Furthermore, we present the structural models from the statistical analysis and the demographics of the participants. Lastly, we conclude with the managerial, the theoretical and the educational implications.

## 2. RESEARCH OBJECTIVES AND HYPOTHESES

The present study presents a comprehensive framework based on the Technology Acceptance Model (TAM) to examine the acceptance of Metaverse technology in the field of higher education among students. Two distinct populations, specifically university students from the Netherlands and Greece, were selected to participate in a survey in order to compare and contrast the results, taking into consideration the cultural differences that exist between these two populations as evidenced by previous research (Hofstede et al., 2005, Hampden-Turner et al., 2020). The unique cultural proclivities of these populations towards exploring communication styles and social etiquette render them intriguing subjects for scholarly investigation in the fields of intercultural communication and cultural studies.

In what follows, we present a detailed comparison of the demographic characteristics and intended behaviors towards new technologies and digital tools related to various aspects of daily life, such as wellbeing, leisure, education, and social life, between the two examined populations. The findings of this comparison reveal noteworthy cultural differences in perspective and acceptance of these new technologies and tools.

Additionally, the survey includes general questions pertaining to students' relationship with new technologies in their daily lives, which, when combined with the TAM-related questions, provide a holistic and comprehensive understanding of students' behaviors and attitudes towards MetaEducation.

Subsequently, we present two distinct frameworks, one for each country, to facilitate comparison and analysis. The generated structural models offer valuable insights for academic teachers, academic ethic committees, policy-makers, and managers in improving existing infrastructures and formulating or adapting future teaching methodologies in the realm of Metaverse technology in higher education.

The present study empirically tests and provides support for several hypotheses related to university students' behavioral intention to use MetaEducation, as well as their attitude, perceived usefulness, and perceived ease of use towards this technology. The following hypotheses are examined:

H1a: University students' attitude positively influences their behavioral intention to use MetaEducation.

H1b: University students' perceived usefulness positively influences their behavioral intention to use MetaEducation.

H1c: University students' perceived ease of use positively influences their behavioral intention to use MetaEducation.

H1d: University students' self-efficacy positively influences their behavioral intention to use MetaEducation.

H1e: University students' subjective norms positively influence their behavioral intention to use MetaEducation.

H2a: University students' perceived usefulness positively influences their attitude toward the use of MetaEducation.

H2b: University students' perceived ease of use positively influences their attitude toward the use of MetaEducation.

H2c: University students' self-efficacy positively influences their attitude toward the use of MetaEducation.

H2d: University students' subjective norms positively influence their attitude toward the use of MetaEducation.

H3a: University students' perceived ease of use positively influences their perceived usefulness to use MetaEducation.

H3b: University students' self-efficacy positively influences their perceived usefulness to use MetaEducation.

H3c: University students' subjective norms positively influence their perceived usefulness to use MetaEducation.

H4a: University students' self-efficacy positively influences their perceived ease of use towards the use of MetaEducation.

H4b: University students' subjective norms positively influence their perceived ease of use towards the use of MetaEducation.

These hypotheses will be tested using statistical analysis of data collected from the study's participants, providing insights into the relationships between various factors and university students' acceptance and intention to use MetaEducation.

## 3. LITERATURE REVIEW

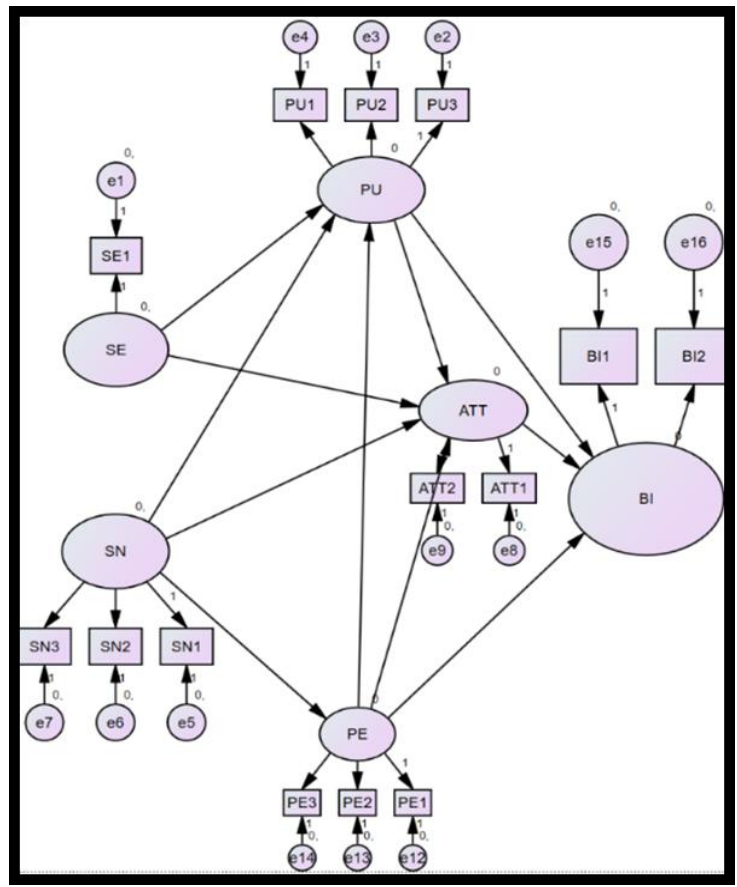

Figure 1. Theoretical Model

The present research uses the well-known TAM, first introduced by Davis (1985) as an extension of Ajzen and Fishbein's Theory of Reasoned Action - TRA (Al-Suqri and Al-Kharusi, 2015). Several studies focusing on education have been already conducted (Scherer et al., 2019, Al-Emran et al., 2018, Weerasinghe and Hindagolla, 2017), but to the best of our knowledge, this is the first one searching the acceptance behaviour of technologies applied in education, related to the Metaverse that compares students from different countries. TAM is a model that explains individuals' intention to accept a certain technology.

Metaverse represents a rather new field of research in science. Even if the term, and what represents, is known for decades, the studies on that matter remain still limited. Despite this limitation, though, researchers understand already that the importance of use of Metaverse technologies in education is crucial and important. The study of Collins (2008) examines the use of Metaverse in education from a future and theoretical perspective. On the other side, the study of Hwang and Chien (2022) examines the subject from an artificial intelligence perspective. Tlili et al. (2022) start their research with an ethical dilemma, whether MetaEducation is a blessing or not. Together with those, several other studies examine the topic from a theoretical perspective, mostly (Singh et al., 2022, Contreras et al., 2022, Suh and Ahn, 2022). Other studies have focused on teachers, exploring the readiness of educators to adopt Metaverse technology for their activities (Lee and Hwang, 2022). Misirlis and Munawar (2022) investigate the utilization of MetaEducation within a specific cultural group, highlighting the presence of low readiness or reluctance towards adopting this technology, supported by empirical evidence. A similar conclusion has been reached by other studies investigating the opinion of students about the potential usefulness of Metaverse and finding a general reluctance to use the technology (Talan and Kalinkara, 2022).

The results in the next paragraphs show several common behaviors but also some differences, mostly because of the different cultural backgrounds of our sample.

## 4. METHODOLOGY

The proposed model with the hypotheses was tested using Structural Equation Modeling (SEM) with maximum likelihood estimation, a technique that examines the covariance structure and relationships between latent variables, accounting for direct, indirect, reciprocal, and misleading causal relationships. A key feature of SEM is its recognition that variables cannot be measured with absolute precision, and therefore includes an error term for measurements. This nuanced approach to measurement is a key advantage of SEM, as it allows for the estimation of latent variables, which are not directly observable, by utilizing multiple indicators that serve as proxies. By accounting for measurement error, SEM enables a more reliable and valid assessment of the underlying constructs being measured, leading to a more robust and comprehensive representation of the intricate relationships among variables. This analytical approach contributes to the robustness and validity of the research findings, enhancing the overall rigor and quality of the study.

In SEM, two models are created: the measurement model and the structural model (one for each country). The measurement model represents the latent constructs using the observed variables, and Confirmatory Factor Analysis (CFA) is employed to verify the factor structure of the observed variables and their underlying latent constructs. This confirms that the latent variables are adequately measured, meeting the standards of measurement.

Figure 1 represents the measurement (theoretical) model to be tested and analyzed. The directed arrows show the relationship between the latent variables and the observed ones. PE and PU can be considered cognitive constructs. Based on our theoretical model, the dataset of our survey was applied to produce the structural models (Figures 2 & 3).

SPSS 21 and SPSS AMOS 21 were used to determine the measurement and structural models. The measurement model included 14 items that described the 6 latent constructs.

Fit indices, such as absolute fit, incremental fit, and comparative fit, were used to assess the goodness of fit of the measurement model, including $x^2$/d.f., non-norm fit index (NNFI), root mean square error of approximation (RMSEA), adjusted goodness of fit index (AGFI), goodness of fit index (GFI), comparative fit index (CFI), and root mean square residual (RMR). The proposed models showed good fit with the collected data, as indicated by the fit indices in Table 1. This allowed for the calculation and evaluation of reliability and validity (convergent and discriminant) of the structural models.

Table 1. The Models' Fit Indices

| Fit Indices | Recommended value | Measurement model | Structural model (the Netherlands) | Structural model (Greece) |
|---|---|---|---|---|
| $x^2$/ d.f. | ≤ 3.00 | 2.17 | 2.13 | 2.14 |
| NNFI | ≥ 0.90 | 0.94 | 0.95 | 0.95 |
| RMSEA | ≤ 0.09 | 0.048 | 0.049 | 0.049 |
| AGFI | ≥ 0.80 | 0.92 | 0.87 | 0.87 |
| GFI | ≥ 0.90 | 0.93 | 0.90 | 0.91 |
| CFI | ≥ 0.90 | 0.92 | .094 | .094 |
| RMR | ≤ 0.05 | 0.046 | 0.049 | 0.049 |

## 5. ANALYSIS OF THE RESULTS

Table 2 summarizes the hypotheses of the present study and the coefficients. With RED color we present the hypotheses not confirmed and with BLUE the confirmed ones. Again, we divide the results in two columns, one for each sample of interest. For those components that not enough data were collected, we leave an empty space.

Table 2. Hypotheses' Paths and Coefficients

| Hypothesis | Path | Coefficient (the Netherlands) | Coefficient (Greece) |
|---|---|---|---|
| H1a | ATT→BI | 0.42 | 0.70 |
| H1b | PU→BI | | |
| H1c | PE→BI | 0.33 | 0.35 |
| H1d | SE→BI | | |
| H1e | SN→BI | | |
| H2a | PU→ATT | | 0.20 |
| H2b | PE→ATT | | 0.35 |
| H2c | SE→ATT | -1.97 | -1.17 |
| H2d | SN→ATT | 2.43 | 1.69 |
| H3a | PE→PU | | |
| H3b | SE→PU | 0.26 | 0.46 |
| H3c | SN→PU | 0.51 | 0.76 |
| H4a | SE→PE | | |
| H4b | SN→PE | 0.91 | 1.0 |

In the analyzed Dutch sample, a significant negative correlation was observed between SE and ATT, whereas SE showed a positive correlation with SN. Similar results were obtained in the Greek sample, indicating consistent findings across both populations. Furthermore, no strong correlations were found between PU and ATT or PU and BI in either sample. Despite meeting acceptable statistical standards, the overall Dutch model exhibited weak explanatory power. Interestingly, perceived ease of use had a minor impact on the final behavior of students, while perceived usefulness did not show any significant effect. In comparison, the Greek sample displayed stronger positive correlations between SN and PU, SE and PU, ATT and BI, and PE and BI. Nevertheless, the overall Greek model also exhibited weak explanatory power, albeit meeting statistical acceptability criteria.

For hypotheses H1b, H1d, H1e, H2a (Dutch sample only), H2b (Dutch sample only), and H3a and H4a, the dataset from both countries did not yield measurable results, precluding any conclusive confirmation or rejection of these hypotheses based on statistical analysis. Notably, the authors' initial expectations of a strong and positive correlation between SE and ATT were contradicted by the observed negative and significant correlation between these two constructs. A potential explanation for this finding is that students with high self-efficacy may be more aware of the difficulties associated with the use of technologies and therefore have a negative attitude towards it.

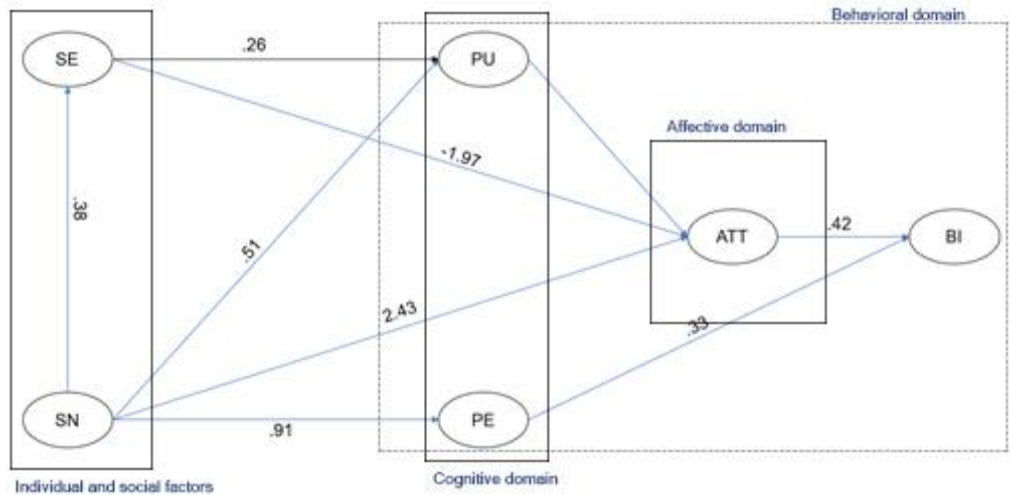

Figure 2. Structural Model – Dutch Dataset

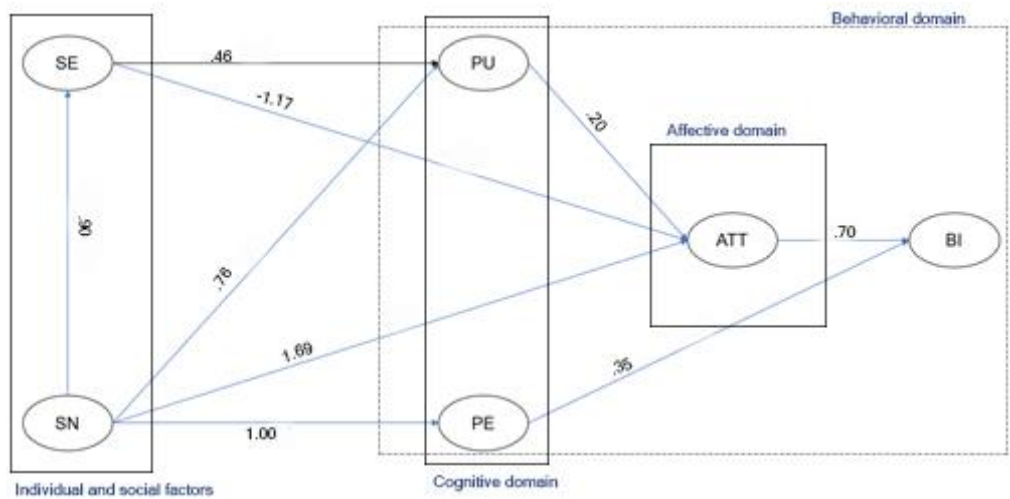

Figure 3. Structural Model – Greek Dataset

## 6. DEMOGRAPHICS AND DATASET

The study garnered a total of 513 valid responses from university students in the Netherlands (n = 285) and Greece (m = 228), with the aim of delving into their relationship with the digital realm and their familiarity with emerging technologies such as the Metaverse. Prior to administering the Technology Acceptance Model (TAM) questionnaire, participants were

25

queried on preliminary aspects to gain deeper insights into their technological engagement. As depicted in Table 2, the gender distribution of the sample was carefully considered and is presented for both countries.

Table 2. Gender Percentages

|  | **The Netherlands (n: 285)** | **Greece (m: 228)** |
| --- | --- | --- |
| **Gender** | Male: 63.4% | Male: 60.5% |
|  | Female: 32.4% | Female: 38.2% |
|  | Other/ prefer not to say: 3.2% | Other/ prefer not to say: 1.3% |

Table 3 provides a breakdown of responses that are not directly linked to TAM, but rather pertain to general technology usage. Remarkably, respondents from Greece exhibited a more positive outlook compared to their counterparts from the Netherlands in terms of the future of education (92.1% vs. 72.4%). Moreover, Greeks expressed a stronger belief in the reliance of their social life and creativity on technology, in contrast to the Dutch population (67.9% vs. 55.6% and 69.3% vs. 42.9%, respectively). Furthermore, Greeks were found to be more inclined to turn to technology for relaxation purposes, surpassing the Dutch respondents by a significant margin (85.6% vs. 66.1%). Interestingly, when it comes to partying, the majority of Greeks (94.7%) preferred physical gatherings, whereas a relatively lower percentage of Dutch respondents (30%) were open to digital parties. A detailed breakdown from the aforementioned statistics and some important questions are listed below:

Table 3. Generic Questions related to Technology Usage

| Question | Dutch sample | Greek sample |
| --- | --- | --- |
| **Do you have an active subscription on at leans one digital platform (Netflix, Amazon, Disney+, etc.)?** | 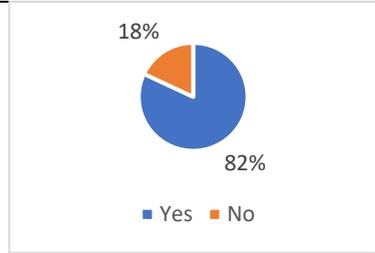 18% / 82% Yes / No | 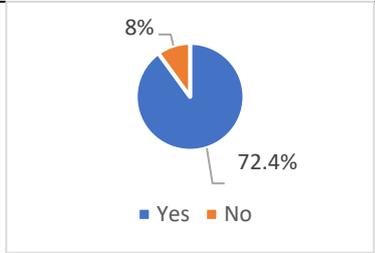 8% / 72.4% Yes / No |
| **Technology is the future of how education will be conducted** | 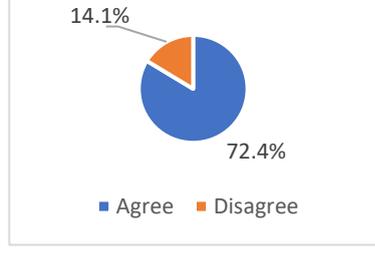 14.1% / 72.4% Agree / Disagree | 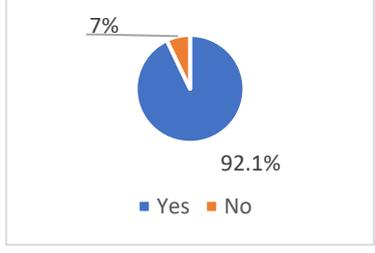 7% / 92.1% Yes / No |

| Question | Dutch sample | Greek sample |
|---|---|---|
| **Technology makes me feel better mentally** | 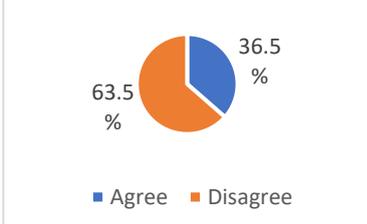 Agree 36.5%, Disagree 63.5% | 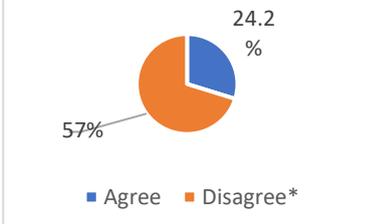 Agree 24.2%, Disagree* 57% |
| **Technology makes me feel better physically:** | 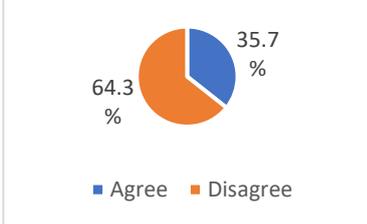 Agree 35.7%, Disagree 64.3% | 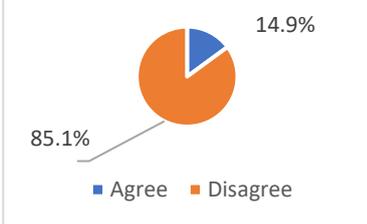 Agree 14.9%, Disagree 85.1% |
| **The future of education is:** | Hybrid: 90.8%<br>Traditional only: 2.3%<br>Digital only: 6.9% | Hybrid: 90.8%<br>Traditional only: <1%<br>Digital only: 8% |
| **I think digital education will benefit the climate crisis by creating more sustainable infrastructures** | 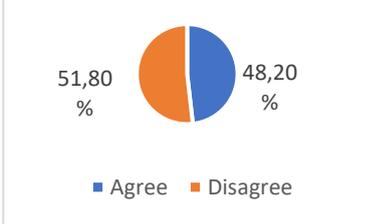 Agree 48,20%, Disagree 51,80% | 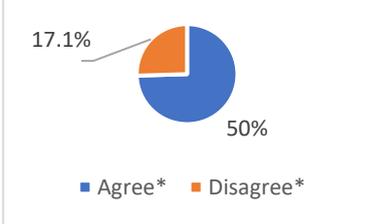 Agree* 50%, Disagree* 17.1% |

\* the total amount of some answers is less than 100%, since some respondents did not fill in this answer.

It is noteworthy that despite being university students who are adept at utilizing various aspects and tools of technology, the respondents exhibit a skeptical stance towards the utilization of the Metaverse in education. This skepticism may stem from multiple factors. Firstly, a lack of familiarity with the concept of the Metaverse in education could contribute to skepticism among the respondents. Previous studies have found that the innovativeness of the academic environment can have an influence on the attitude towards technology adoption (Almarzouqi et al., 2022). While students may be proficient in using technology in their daily lives, the concept of the Metaverse in an educational context may still be novel to them, leading to a limited understanding of its workings and its potential for effective integration into educational settings. Secondly, concerns about privacy and security may also contribute to skepticism towards the use of the Metaverse in education. The creation of virtual environments for user interaction with digital content and other users raises concerns about how personal data and information may be used, as well as the potential for cyber threats and malicious activities. University students may express apprehension about the safeguarding of their privacy and security in such virtual environments.

Pedagogical concerns may also contribute to skepticism among students. They may question the efficacy of virtual environments in replicating the rich learning experiences of real-world interactions, such as face-to-face discussions, group projects, and hands-on learning. Doubts may arise about the pedagogical effectiveness of the Metaverse in facilitating meaningful educational experiences. Ethical concerns could also be a source of skepticism among students. Issues related to inclusivity, diversity, and accessibility may be questioned in the context of the Metaverse in education. New models of education toward new technologies, sustainability and global citizenship should take the aforementioned concerns into consideration (Misirlis, 2023). Students may express apprehension about the potential for perpetuating biases and inequalities in a digital space, and the ethical aspects of such implications. Additionally, technical challenges associated with the Metaverse may also contribute to skepticism among students. As the Metaverse is still a developing technology, issues such as connectivity, hardware requirements, and software compatibility may pose challenges that need to be addressed. Such technical challenges could contribute to doubts about the feasibility and practicality of implementing the Metaverse in an educational context.

Further research and exploration are necessary to address these concerns and to fully understand the potential of the Metaverse in the field of education. The findings from this study shed light on the contrasting perspectives and attitudes towards technology and its impact on various aspects of life among university students in the Netherlands and Greece. The results highlight the need for further investigation and understanding of cultural and contextual factors that shape individuals' perceptions and behaviors towards technology in different regions.

# REFERENCES

Al-Emran, M., Mezhuyev, V. & Kamaludin, A. (2018). Technology Acceptance Model in M-learning context: A systematic review. *Computers & Education,* 125**,** 389-412.

Al-Suqri, M. N. & Al-Kharusi, R. M. (2015). Ajzen and Fishbein's theory of reasoned action (TRA)(1980). *Information seeking behavior and technology adoption: Theories and trends.* IGI Global.

Almarzouqi, A., Aburayya, A. & Salloum, S. A. (2022). Prediction of user's intention to use metaverse system in medical education: A hybrid SEM-ML learning approach. *IEEE access,* 10**,** 43421-43434.

Collins, C. (2008). Looking to the future: Higher education in the Metaverse. *Educause Review,* 43**,** 51-63.

Contreras, G. S., González, A. H., Fernández, M. I. S. & Martínez, C. B. (2022). The Importance of the Application of the Metaverse in Education. *Modern Applied Science,* 16**,** 1-34.

Davis, F. D. (1985). *A technology acceptance model for empirically testing new end-user information systems: Theory and results.* Massachusetts Institute of Technology.

Hampden-Turner, C., Trompenaars, F. & Hampden-Turner, C. (2020). *Riding the waves of culture: Understanding diversity in global business*, Hachette UK.

Hofstede, G., Hofstede, G. J. & Minkov, M. (2005). *Cultures and organizations: Software of the mind*, Mcgraw-hill New York.

Hwang, G.-J. & Chien, S.-Y. (2022). Definition, roles, and potential research issues of the metaverse in education: An artificial intelligence perspective. *Computers and Education: Artificial Intelligence***,** 100082.

Kaddoura, S. & Al Husseiny, F. (2023). The rising trend of Metaverse in education: challenges, opportunities, and ethical considerations. *PeerJ Computer Science,* 9**,** e1252.

Ko, S. Y., Chung, H. K., Kim, J.-I. & Shin, Y. (2021). A study on the typology and advancement of cultural leisure-based metaverse. *KIPS Transactions on Software and Data Engineering,* 10**,** 331-338.

Lee, H. & Hwang, Y. (2022). Technology-enhanced education through VR-making and metaverse-linking to foster teacher readiness and sustainable learning. *Sustainability,* 14**,** 4786.

Misirlis, N. (2023). I Am a Global Citizen. Or Am I Not? International Business Schools' Students and Global Citizenship Unified Framework & A Scoping Literature Review of The Last Decade (2013-2022) *International Conference - The Future of Education 13th Edition 2023.* Florence-Italy.

Misirlis, N., Elshof, M. & Vlachopoulou, M. (2021). Modeling Facebook users' behavior towards the use of pages related to healthy diet and sport activities. *Journal of Tourism, Heritage & Services Marketing (JTHSM),* 7**,** 49-57.

Misirlis, N. & Munawar, H. B. (2022). An analysis of the technology acceptance model in understanding university students' behavioral intention to use metaverse technologies. 12th International Conference: The Future of Education, 2022. 159-163.

Park, S. Y. (2009). An analysis of the technology acceptance model in understanding university students' behavioral intention to use e-learning. *Journal of Educational Technology & Society,* 12**,** 150-162.

Scherer, R., Siddiq, F. & Tondeur, J. (2019). The technology acceptance model (TAM): A meta-analytic structural equation modeling approach to explaining teachers' adoption of digital technology in education. *Computers & Education,* 128**,** 13-35.

Singh, J., Malhotra, M. & Sharma, N. (2022). Metaverse in Education: An Overview. *Applying Metalytics to Measure Customer Experience in the Metaverse***,** 135-142.

Suh, W. & Ahn, S. (2022). Utilizing the Metaverse for Learner-Centered Constructivist Education in the Post-Pandemic Era: An Analysis of Elementary School Students. *Journal of Intelligence,* 10**,** 17.

Talan, T. & Kalinkara, Y. (2022). Students' Opinions about the Educational Use of the Metaverse. *International Journal of Technology in Education and Science,* 6**,** 333-346.

Terdiman, D. (2007). *The entrepreneur's guide to Second Life: Making money in the metaverse*, John Wiley & Sons.

Tlili, A., Huang, R., Shehata, B., Liu, D., Zhao, J., Metwally, A. H. S., Wang, H., Denden, M., Bozkurt, A. & Lee, L.-H. (2022). Is Metaverse in education a blessing or a curse: a combined content and bibliometric analysis. *Smart Learning Environments,* 9**,** 1-31.

Weerasinghe, S. & Hindagolla, M. (2017). Technology acceptance model in the domains of LIS and education: A review of selected literature. *Library Philosophy and Practice***,** 1-27.